\documentclass[%
reprint,
superscriptaddress,
 amsmath,
 amssymb,
 aip,jcp,
]{revtex4-2}

\usepackage{graphicx}
\usepackage{animate}
\usepackage{physics}
\usepackage{dcolumn}
\usepackage{bm}
\usepackage{subfigure}
\usepackage[dvipsnames]{xcolor}
	
\usepackage{color, colortbl}

\usepackage[pdftex]{hyperref}
	\hypersetup{
	colorlinks=true,       
    	linkcolor=red,          
   	citecolor=blue,        
    	filecolor=magenta,      
    	urlcolor=cyan           
	}

\usepackage{graphicx}
\usepackage{dcolumn}
\usepackage{bm}
\usepackage{hyperref}
\usepackage{tikz}

\definecolor{Gray}{gray}{0.9}
\definecolor{LightCyan}{rgb}{0.88,1,1}
\definecolor{hl}{rgb}{1,0,0}
\definecolor{marker}{RGB}{242,224,48}

\begin{document}

\title{A quantum analog of Huygen's clock: 
noise-induced synchronization}

\author{Bhavay Tyagi}
\affiliation{Department of Physics, University of Houston, Houston, Texas 77204, United~States.}

\author{Hao Li}
\affiliation{
Department of Chemistry, University of Houston, Houston, Texas 77204, United~States.}
\affiliation{Institut Courtois \& D\'epartement de physique, Universit\'e de Montr\'eal, Caisse Postale 6128, Succursale centre-ville, Montr\'eal, Qu\'ebec H3T~3J7, Canada}

\author{Eric~R.~Bittner}
\email{ebittner@central.uh.edu}
\affiliation{Department of Physics,
University of Houston, Houston, Texas 77204, United~States.}

\author{Andrei Piryatinski}
\affiliation{Theoretical Division, Los Alamos National Laboratory, Los Alamos, New Mexico 87545, United States}

 \author{Carlos~Silva-Acu\~na}
 \email{carlos.silva@umontreal.ca}
 \affiliation{Institut Courtois \& D\'epartement de physique, Universit\'e de Montr\'eal, Caisse Postale 6128, Succursale centre-ville, Montr\'eal, Qu\'ebec H3T~3J7, Canada}


\date{\today}
\begin{abstract}
We propose a quantum analogue of the Huygens clock, in which the phases of two spins achieve synchronization through their interaction with a shared environment. The environment functions analogously to the escapement mechanism in a mechanical clock, regulating the gear train and permitting the advancement of timing in discrete intervals. In our proposed model, the relative phase of the two spins become synchronized through interaction with a mutual, correlated, environment. We show that for a system of qubits, several arguments can be made that significantly reduce the cardinality of the set of allowed measurements and, hence, the complexity of the problem. We present a numerically efficient method to calculate the degree of quantumness that exists in the correlations of our final density matrix. This method also provides a tight upper bound for when the system is described by rank-3 and rank-4 density matrices.
\end{abstract}

\maketitle

\section{Introduction}

At the close of the 17th century, Christiaan Huygens observed that two identical pendulum clocks, when weakly coupled through a substantial beam, rapidly synchronized with identical periods and amplitudes, yet with the pendula oscillating in antiphasing directions. \cite{Huygens:1669aa,Huygens:1673aa,Ellicott:1739aa,Ellicott:1739ab,Airy:1830aa,Newton:1833aa,Korteweg:1906aa,huygens1916oeuvres}
He introduced his advances in pendulum clock technology to the Royal Society with the intention of addressing the "longitude problem," which was pivotal for ascertaining a ship's precise position at sea.
Initially, Huygens attributed the synchronization of pendulum clocks to air currents. However, subsequent experiments led him to surmise that the weak coupling via the beam instigated this antiphase synchronization. Currently, oscillator synchronization is recognized as a foundational concept in nonlinear science, encompassing both the classical and quantum domains.\cite{Strogatz:1993aa,Pikovsky:2001aa,Bennett:2002aa,Goychuk:2006aa,Oliveira:2015aa,Pena-Ramirez:2016aa}
 
In this paper, we propose a quantum analogue of the Huygens clock, wherein the phases of two spins achieve synchronization through their interaction with a shared environment. In our proposed model, the environment functions analogously to the escape mechanism in a mechanical clock, regulating the gear train and permitting the advancement of time in discrete intervals. The escapement also imparts a nominal amount of kinetic energy to the pendulum by releasing potential energy stored in a coiled spring or a series of weights. In our framework, escapement is represented by a stochastic term that induces the transition from $|0\rangle$ to $|1\rangle$ at a specified rate. 
\begin{figure}
    \centering
\includegraphics[width=\columnwidth]{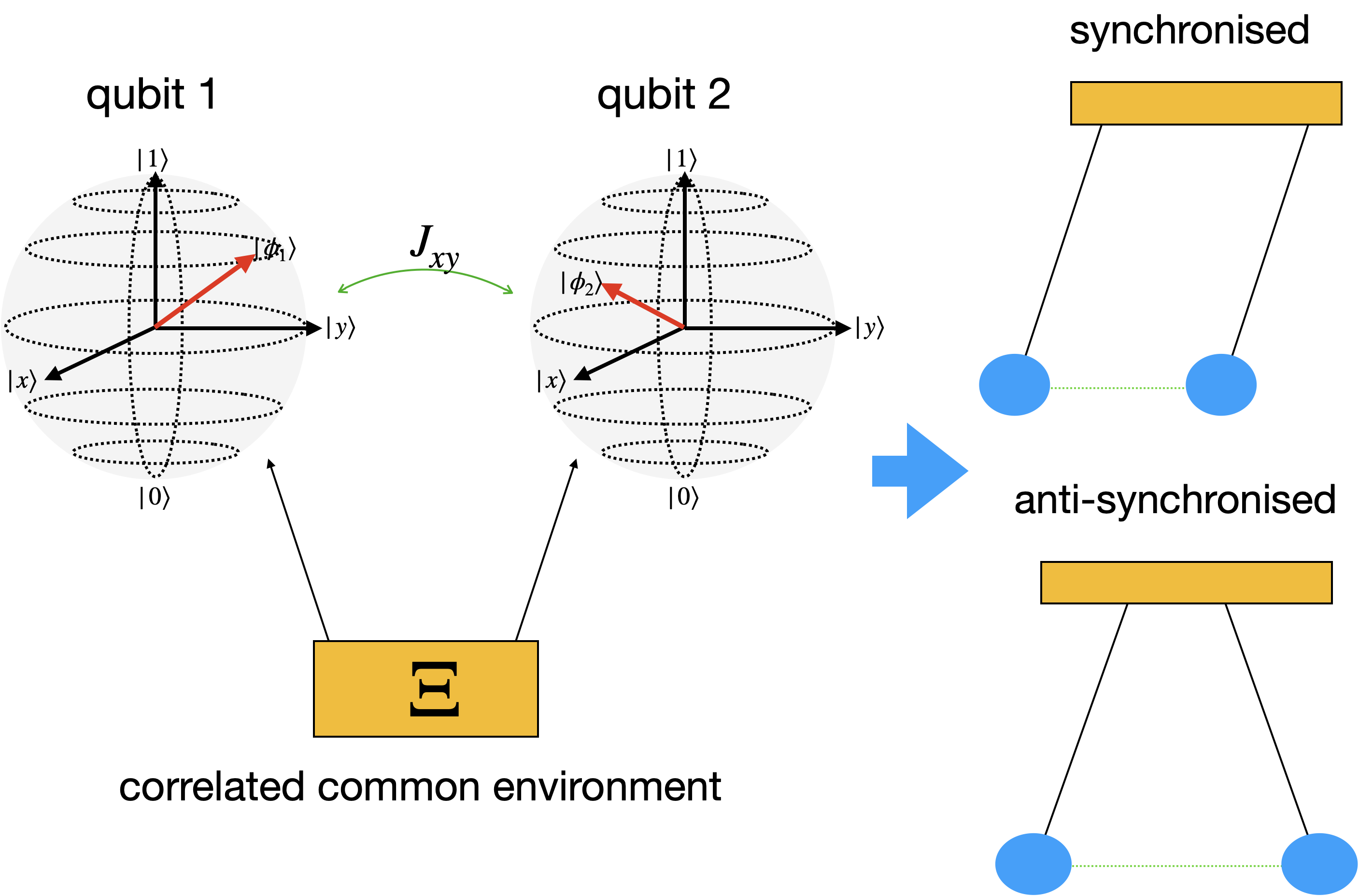}
    \caption{Model for quantum ``clocks'' consisting
    of two weakly coupled qubits coupled to local environments with known correlations.}
    \label{fig:1}
\end{figure}
Our results and analysis
indicate that both synchronization and antisynchronization can be achieved
within this model, depending upon the degree of correlation between the escapement mechanisms acting on the two
states. Our results have implications in terms of
developing quantum technologies
that are robust against
environmental effects such as
decoherence, as well as in
understanding the underlying
mechanisms for the transport
of quantum excitations in 
photosynthetic systems
in which observed long-lived exciton coherences are linked to vibronic correlation effects.  The enhanced excitonic coherence has a profound influence on the enhancement of exciton transport and delocalization. \cite{engel2007evidence,panitchayangkoon2010long,chenu2015coherence,scholes2017using,Zhu:2024aa,Du:2021aa}

\section{Theoretical model}
The scenario we envision
is figuratively sketched
in Fig.~\ref{fig:1} 
whereby the two qubits
are directly coupled through
an exchange term $J_{xy}$ and indirectly through correlations within their local environments
indicated by $\Xi$.
A suitable Hamiltonian corresponding to this scenario is given by 
\begin{align}
    H &= \frac{\Delta}{2}(\sigma^z_1+\sigma^z_2)
    +
    \frac{\tau}{2}
    \left( \sigma^x_1+ \sigma^x_2\right)
    +
    J_{xy} (\sigma^+_1\sigma^-_2
    + \sigma^+_2\sigma^-_1)
    \label{eq:hamiltonian}
\end{align}
Similar models have been 
studied by us and others
\cite{Brox_2012,Eneriz:2019aa,Bittner:2024aa}.
In this model, each
local site has a characteristic
Rabi frequency $\hbar\tau$
that produces a coherent
oscillation between
local states $|0_i\rangle$ and $|1_i\rangle$.  The qubits
are coupled via $J_{xy}$, which
allows the transfer of
a quantum from one qubit to the
other. $\Delta$ sets the energy gap between $|0_i\rangle$ and $|1_i\rangle$
for the decoupled spins.

Decoherence and dissipation
are introduced into our model via the Lindblad term. However, 
we also introduce correlation between
the local environments
by first writing the system/bath 
interaction in the form.
\begin{align}
    H_{s/b} = \sum_{i=1,2}
    \mathbf{\sigma}_i E_i(t)
\end{align}
where each $\sigma_i$ is a local 
spin operator and $E_i(t)$ is a 
stochastic term. 
For our purposes,  we take the
driving terms to be correlated Ornstein-Uhlenbeck 
processes with
\begin{align}
    d\mathbf{E} = -\mathbf{A}\cdot \mathbf{E} dt + \mathbf{B}\cdot d\mathbf{W}
    \label{eq:3}
\end{align} 
with $\mathbf{\Xi}dt\delta(t-t') = d\mathbf{W}\otimes d\mathbf{W}$
as the correlation matrix,
\begin{align}
    \Xi = \left(
    \begin{array}{cc}
        1 &\xi  \\
        \xi & 1 
    \end{array}
    \right).
\end{align}
Working out 
the covariance and cross-covariance terms, we obtain the 
spectral density for the 
noise as 
\begin{align}
    \mathbf{J}(\omega) 
    = \frac{1}{2\pi}(\mathbf{A}+i\omega)^{-1}\cdot\mathbf{B}\cdot\mathbf{\Xi}\cdot \mathbf{B}^T\cdot (\mathbf{A}-i\omega)^{-1}
\end{align}
which is a $2\times 2$ matrix. 
We bring this into diagonal form by defining a transformation matrix $T$ that diagonalizes the $\Xi$ matrix (assuming that both
$\mathbf{A}$ and $\mathbf{B}$
are diagonal to begin with).
This has the effect of transforming the
two correlated Wiener processes
into a pair of uncorrelated processes
that satisfies the Ito
identity $d\tilde W_\alpha(t)d\tilde W_\beta(t') = \delta(t-t')dt$\cite{Bittner:2024aa}.

Following this, we can write the system/bath interaction as 
\begin{align}
H_{s/b} = \sum_\alpha \tilde \sigma_\alpha^{(n)} \tilde E_\alpha^{(n)}(t).
\end{align}
where we transform the system
operators as 
\begin{align}
    \tilde\sigma_\alpha^{(n)} = \sum_{i=1,2}
    T_{\alpha,i}\sigma_i^{(n)}
\end{align}
The usual series of
assumptions allow us to
include the stochastic
terms in the form of a series of
quantum jump processes described 
by Lindblad operators of the form
\begin{align}
    c^{(n)}_S &=\sqrt{ \gamma (1+\xi)}
    \frac{1}{\sqrt{2}}
    (\sigma_1+\sigma_2)\\
    c^{(n)}_A&= \sqrt{ \gamma (1-\xi)}
    \frac{1}{\sqrt{2}}
    (\sigma_1-\sigma_2)
\end{align}
corresponding to symmetric and
anti-symmetric system/bath interactions and 
$\gamma$ gives the rate for this 
interaction. Note, that $\sigma_1$ and
$\sigma_2$ can be any operator acting 
on spin 1 or spin 2, respectively. 

From this we write the Liouville-von Neumann equation for the system-density matrix as follows.
\begin{align}
    \partial_t \rho
    = \frac{1}{i\hbar}
    [H_{sys},\rho] 
    +\left( 
    {\cal D}_S(\rho)+ {\cal D}_A(\rho)
    \right)
    \label{eq:lvn}
\end{align}
where each ${\cal D}_k(\rho)$ is 
the Lindblad dissipator
\begin{align}
    {\cal D}_k(\rho)
     = c_k \rho c_k^\dagger
     - \frac{1}{2}
     \{c_k^\dagger c_k, \rho\}.
\end{align}
where $c_k$ are the Lindblad
operators.
The two dissipators in Eq.~\ref{eq:lvn}
can be combined in terms of site-local terms and cross-terms
\begin{align}
\left( 
    {\cal D}_S(\rho)+ {\cal D}_A(\rho)
    \right)
    = {\cal D}_1(\rho) + {\cal D}_2(\rho)
    + \xi{\cal D}_{12}(\rho)
\end{align}
where
\begin{align}
    {\cal D}_{12}(\rho)
    &=\gamma
    \left(
\sigma_1\rho\sigma^\dagger_2
+\sigma_2\rho\sigma^\dagger_1) 
-
    \frac{1}{2}
    \{
\sigma^{\dagger}_1\sigma_2+\sigma^{\dagger}_2\sigma_1,\rho
    \}
    \right)
\end{align}
The crucial parameter in this is $\xi$, which gives the
correlation between the two local 
environments.
As we shall show, this cross-term has the
effect of creating a correlation between two spins even
when the exchange coupling is $J_{xy} = 0$.

As mentioned above, a key feature of a pendulum clock is the
escapement mechanism, which provides a
small ``kick'' to the pendulum. We introduce this into our model
using the $\sigma_i^+$ operators so
that the escapement provides a stochastic
driving field on the pair of qubits
inducing transitions from the $|0_i\rangle
\to |1_i\rangle$ state at a rate specified by $\gamma$.  
 As one of the
qubits cycles through its $|0_i\rangle$
state over the course of its normal unitary-time evolution, it can be
kicked incoherently to the 
$|1_i\rangle$ state by the interaction
with the environment. If the two qubits interact
with uncorrelated baths ($\xi = 0$), 
the two kick processes are
independent and no phase synchronisation is expected. 
However, the ${\cal D}_{12}(\rho)$ dissipator
introduces cross-terms 
such that the excitation of
one spin is contingent upon the
state of the other spin. 
This induces coherence synchronisation
in $\rho$ by enhancing or suppressing the terms
in-phase or out-of-phase depending
on whether the ``kicks'' are correlated or
anti-correlated.

\begin{figure*}
    \centering
    \subfigure[]{\includegraphics[width=0.24\linewidth]{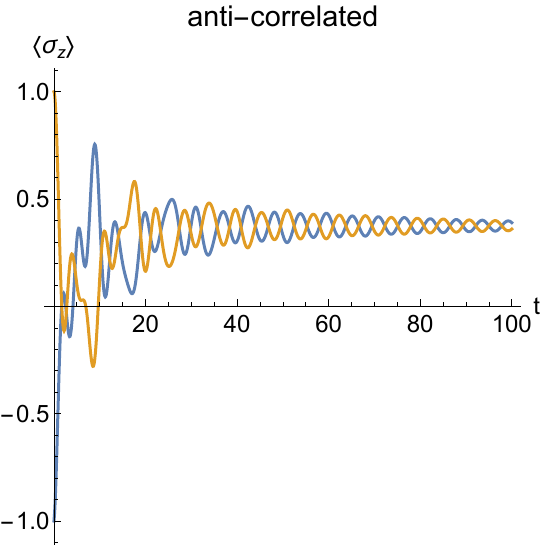}}
    \subfigure[]{\includegraphics[width=0.24\linewidth]{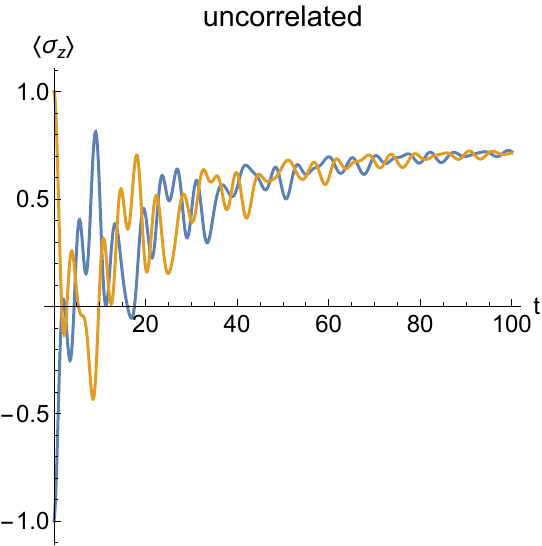}}
      \subfigure[]{ \includegraphics[width=0.24\linewidth]{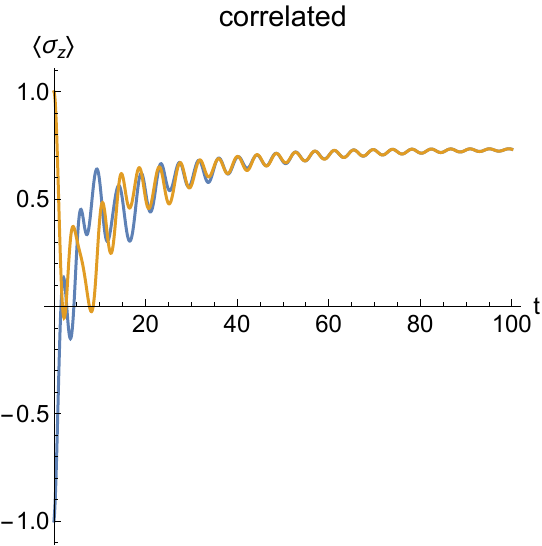}}
     \\ 
      \subfigure[]{ \includegraphics[width=0.24\linewidth]{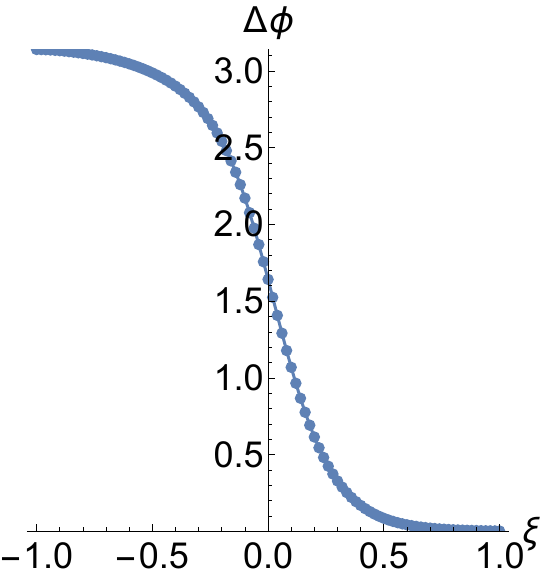}}
      \subfigure[]{ \includegraphics[width=0.24\linewidth]{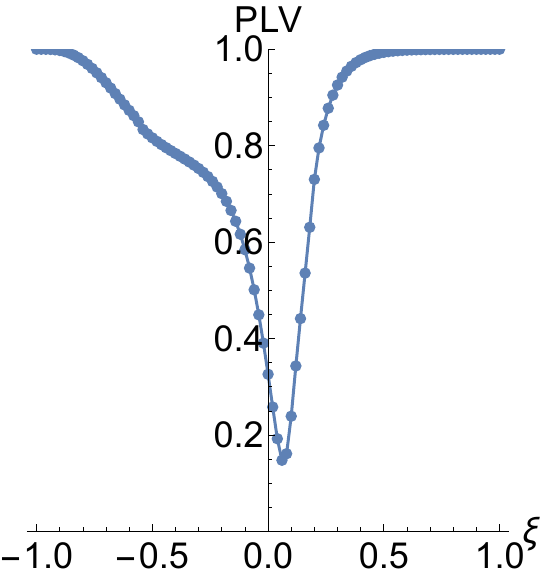}}
    \caption{(a-c) Expectation values 
    of $\langle \sigma_1^z(t)\rangle$ (blue)
    and $\langle \sigma_2^z(t)\rangle$(orange)
    for 2 qubit system subject to various degrees of bath correlation as described
    by Eq.~\ref{eq:hamiltonian}
    with parameters: $\Delta=1$, $\tau = 1$, 
 $J_{xy}=0.25$, and $\gamma=0.05$
in reduced units. (a) Anti-correlated ($\xi = -1$); (b)Uncorrelated ($\xi = 0$); (c) Correlated ($\xi = +1$).
(d,e)Asymptotic phase shift and phase-locking value vs. $\xi$.}
    \label{fig:2}
\end{figure*}
\section{Synchronization Dynamics}
\label{syncdynamics}
To study this effect, we
performed numerical simulations of the
system using the QuTip (v5.0) quantum
simulation package\cite{JOHANSSON20121760,JOHANSSON20131234}, and the representative results of our simulations are
shown in Fig.~\ref{fig:2}. 
In the example given here, we
set $\Delta=1$, $\tau = 1$, 
 $J_{xy}=0.25$, and $\gamma=0.05$
in reduced units. 
In each case, the initial state
is set as $|\psi_0\rangle = |1_10_2\rangle$.
In Fig.~\ref{fig:2} we show the
expectation value of $\langle \sigma_1^z\rangle$ in blue and
$\langle\sigma_2^z\rangle$ in orange
when the environments are (a)
anticorrelated with $\xi =-1$, (b)
uncorrelated with $\xi = 0$, and
(c) correlated with $\xi = +1$.
At long time, the two
qubits become phase-synchronized
depending upon the correlation. 

We quantify this in Fig.~\ref{fig:2}(d) by comparing the asymptotic phase difference between the two qubits
as a function of the correlation parameter. 
For this, we extracted the
last 25\% of the time series for $\langle\sigma_1^z\rangle$ 
and $\langle\sigma_1^z\rangle$ 
subtracted off the baseline
and used the discrete Hilbert transform
to generate the analytic signal for
the two qubits. 
Both the $\Delta \phi$ phase shift and the phase locking value provide useful order parameters
for quantifying the synchronization
between observables. 
In the Supplementary
Information, we include a series
of movies showing the full dynamics
projected onto a common Bloch sphere. 
We conclude from the dynamics that
synchronization or antisynchronization can occur depending
upon the correlation within the
driving forces acting on each qubit. 

At long time, the system 
converges to a steady state
that also depends upon the 
degree of correlation within the
environment since
\begin{align}
    \frac{i}{i\hbar}
    [H_{sys},\rho_{ss}] 
    +
   {\cal D}_1(\rho_{ss}) + {\cal D}_2(\rho_{ss})
    =- \xi{\cal D}_{12}(\rho_{ss}).
\end{align}
Consequently, it is not obvious \textit{a priori} whether $\rho_{ss}$
represents a mixed state
in which $\rho_{ss} = \rho_{A}\otimes\rho_{B}$
or represents an entanglement
between systems A and B.
Furthermore, the question remains as
to whether or not the synchronization
and entanglement between the
two qubits is of quantum or classical
origin.

 A useful way to quantify entanglement in the quantum state is the {\it entanglement entropy} of the state. Given a bipartite system $\mathcal{H}_{AB} = \mathcal{H}_A \otimes \mathcal{H}_B$, the
reduced density matrix describing subregion A of the system is given by tracing over subregion B, 
\begin{equation}
    \rho_A = tr_{B} \ket{\psi}\bra{\psi} 
\end{equation}
where $\ket{\psi} \equiv \ket{\psi}_{AB}$ is the wavefunction of the whole system. The entropy of entanglement of subregion A is defined as the von Neumann entropy (vNE) of the reduced density matrix describing that subregion
\begin{equation}
    S(\rho_A) = -tr \rho_A \ln \rho_A\;.
\end{equation}
This quantity is exactly zero when we know everything about the state and is equal to 
$\log d$ when the state is maximally mixed, i.e., $\ket{\psi}= \mathbb{I}/d$. In order to distinguish between two quantum states we look at the {\em relative entropy} between them,
\begin{align}
S(\rho || \sigma)=tr \left(\rho \ln \rho - \rho \ln \sigma\right) \geq 0.
\end{align}
One can think of this as the amount of information about the distance between the two states. The bound is saturated if and only if $\rho = \sigma$. In a bipartite system, the density matrix representing an arbitrary preparation of the entire system is $\rho_{AB}$. In general $\rho_{AB} \neq \rho_A \otimes \rho_B$. We can then apply the above notion of relative entropy and see how $\rho_{AB}$ and $\rho_A \otimes \rho_B$ differ in terms of the information they contain. This relative entropy is defined
as the {\em mutual information} as given by
\begin{equation}
\begin{split}
    \mathcal{I}_{\rho_{AB}}&(A:B) = S(\rho_{AB}||\rho_A \otimes \rho_B)\\
    &= -S(\rho_{AB}) - tr(\rho_{AB}\ln \rho_A) - tr(\rho_{AB}\ln \rho_B)\\
    &= S(\rho_A) + S(\rho_B) - S(\rho_{AB})
\end{split}
\end{equation}

\begin{figure*}
    \centering
    \subfigure[]{
    \includegraphics[width=0.3\linewidth]{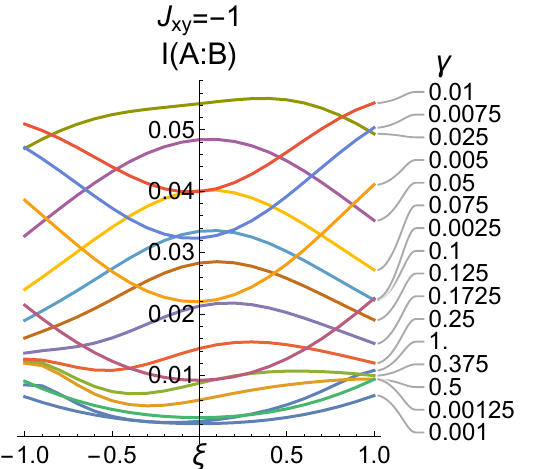}}
    \subfigure[]{
    \includegraphics[width=0.3\linewidth]{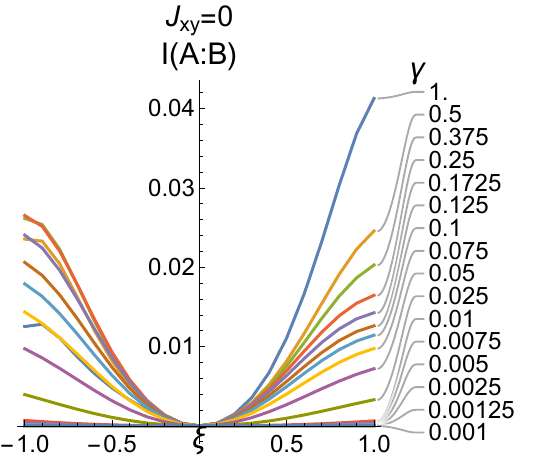}}
    \subfigure[]{
    \includegraphics[width=0.3\linewidth]{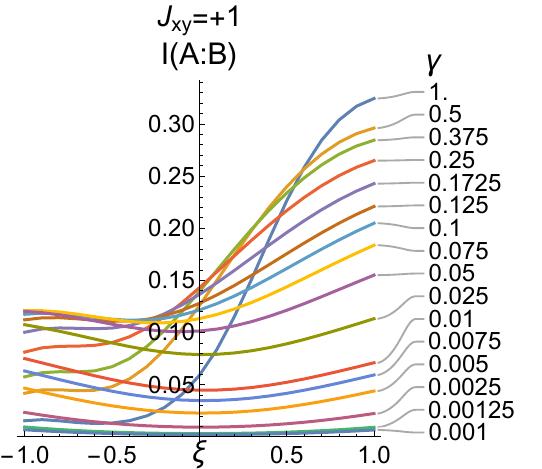}}
    \caption{Mutual Entropy for $J_{xy}=0$
    vs. correlation parameter and coupling 
    constant $\gamma$.}
    \label{fig:3}
\end{figure*}

In Fig.~\ref{fig:3}, we show the mutual
information for steady-state solutions.
as a function of the correlation parameter 
$\xi$ and coupling constant $\gamma$.
For $J_{xy}=0$ and $\xi = 0$, the two spins are completely decoupled, and there is no
mutual information between the two subsystems.
However, for $\xi\ne 0$, correlations 
in the environment lead to increasing degrees
of entanglement even though there is no 
direct coupling between the two spins.  
Increasing $J_{xy}$, leads to increased 
entanglement. 

Mutual information also quantifies the total number of correlations in a given state. These correlations can be classical or quantum.
We know that if $\rho_{AB}$ is a pure state, then the vNE $S(\rho_{AB})$ is zero and the entropies of the subsystems are equal to each other. Therefore, mutual information is
\begin{align}
\mathcal{I}_{\rho_{AB}}(A:B)=
2 S(\rho_A)=2 S(\rho_B).
\end{align}

\begin{figure*}
    \centering
    \subfigure[]{\includegraphics[width=0.3\linewidth]{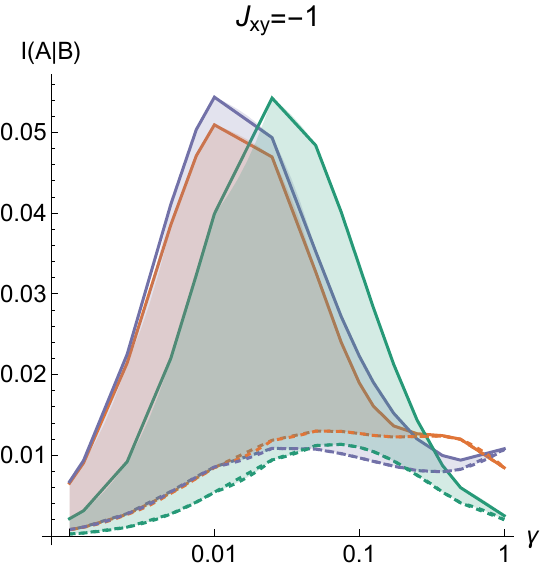}}
    \subfigure[]{\includegraphics[width=0.3\linewidth]{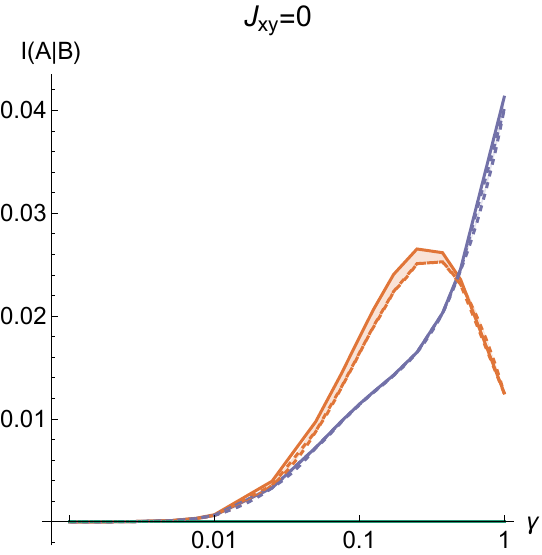}}
    \subfigure[]{\includegraphics[width=0.3\linewidth]{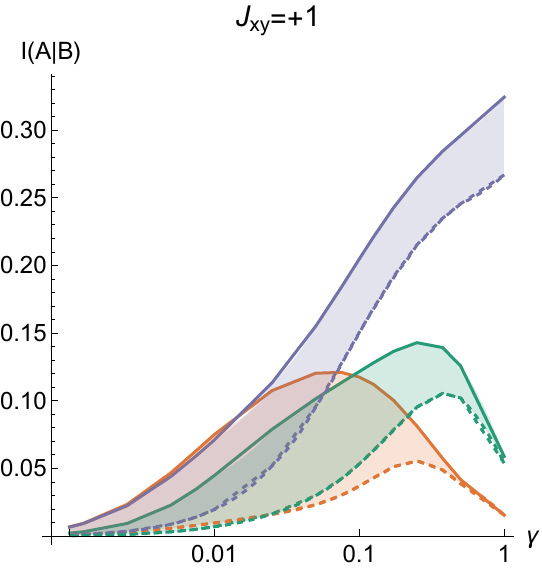}}
    \caption{Mutual information
    at  $J_{xy}=-1$ (a), $J_{xy}=0$ (b), and
    $J_{xy}=+1$ (c) versus the 
    coupling term $\gamma$ for 
    correlated (blue), anti-correlated (orange), 
    and uncorrelated (green) environments.
    The dashed lines represent the contribution due entirely to the \textit{quantum} correlation effects, $D(A:B)$ per equation \ref{eq:disco}.
    The shaded 
    areas denote the 
    difference between the total mutual information and 
    our lower-bound estimate 
    of the ``quantum-ness'' and represents the maximum amount of classical correlation. }
    \label{fig:4}
\end{figure*}
For pure states, all existing correlations are truly quantum and represent entanglement. When dealing with quantum systems, the type of measurement plays an important role since it has a backaction on the system's state. For a bipartite system, we can learn something about subsystem A by making some measurements on subsystem B. Consider a Positive Operator-Valued measurement (POVM) $\Pi_k^B$ that satisfies $\sum_k (\Pi_k^B)^\dagger \Pi_k^B = \mathbb{I}$ and $\bra{\psi}\Pi_k^B\ket{\psi} \geq 0$. We start our system in some arbitrary $\rho_{AB}$ and perform this measurement in subsystem B. We know that the probability of the k-th outcome is given by 
\begin{equation}
    p_k^B = tr \left[(\mathbb{I}_A \otimes \Pi_k^B)\rho_{AB}(\mathbb{I} \otimes \Pi_k^B)^{\dagger}\right]
\end{equation}
and the post-measurement state is 
\begin{equation}
    \rho_{AB|k} = \frac{(\mathbb{I}_A \otimes \Pi_k^B)\rho_{AB}(\mathbb{I} \otimes \Pi_k^B)^{\dagger}}{p_k^B}.
\end{equation}
This allows us to compute the {\em conditional entropy}
\begin{equation}
    S_{\Pi}(A|B)= \sum_k p_k^B S(\rho_{A|k})
\end{equation}
where $S(\rho_{A|k})$ is the reduced density matrix describing subsystem A, given that we get the outcome $k$ when we perform a measurement on subsystem B. In other words, this is the lack of information about state $\rho_A$ when we know the outcomes of our measurement in subsystem B. Note that the conditional entropy depends non-trivially on the choice of measurement. This allows us to filter out the classical part of the total mutual information. It is given by 
\begin{equation}
\begin{split}
    \mathcal{J}_{\Pi} (A|B) &= S(\rho_A) - S_{\Pi}(A|B)\\
    &= S(\rho_A) - \sum_k p_k^B S(\rho_{A|k}).
\end{split}
\end{equation}
In essence, this quantity tells us how much we can learn about subsystem A by performing some local measurements in subsystem B based on the classical correlations that exist between the two subsystems\cite{HendersonVedral:2001}. The difference between the total mutual information $\mathcal{I}(A:B)$ and this classical part is known as the {\em quantum discord}$\;$\cite{OllivierZurek:2002}:
\begin{equation}
    \mathcal{D}_{\Pi}(A|B)= \mathcal{I}(A:B) - \mathcal{J}_{\Pi}(A|B).
\end{equation}
and it quantifies the amount of remaining shared information between subsystems once we gain some new information by measuring subsystem B with our specific choice of the set of measurements $\Pi_k^B$. This provides information about the true quantum correlations that exist between subsystems. Intuitively, this lack of agreement makes sense because measurements play a ``special" role in quantum mechanics and may not give a complete picture. Note that this quantity is not symmetric \textit{i.e.}, $\mathcal{D}(A|B) \neq \mathcal{D}(B|A)$. To make this quantity independent of the choice of measurements, we can try to maximize the classical correlations over all possible POVMs
\begin{align}
\mathcal{D}(A|B)&=\mathcal{I}(A:B)- \max_{\{\Pi_k^B\}}\mathcal{J}_{\{\Pi_k^B\}}(A|B)
\nonumber\\
&= \mathcal{I}(A:B)-S(\rho_A)+ \min_{\{\Pi_k^B\}}\sum_k p_k^B S(\rho_A|k)
\nonumber \\
&=S(\rho_B)-S(\rho_{AB})+\min_{\{\Pi_k^B\}}\sum_k p_k^B S(\rho_A|k)
\end{align}
Therefore,
\begin{equation}
    \mathcal{D}(A|B) = \min_{\{\Pi_k^B\}} \mathcal{D}_{\Pi}(A|B)
\end{equation}
It is known that outcomes expected from local measurements do not completely recover all correlations, and therefore, the discord is a non-zero quantity for such post-measurement states. This minimising procedure is highly nontrivial because infinitely many POVMs exist that differ by a simple change of basis. In addition, finding a measurement that can be made that could maximise $\mathcal{J}(A|B)$ is not straightforward or practical
for anything but a simple system. 
\cite{Eneriz:2019aa,Mohamed:2020aa,Foti:2021aa}

With this in mind, we define the
\textit{classical mutual information} by taking just the
diagonal elements of the density matrix.
\begin{align}
    \tilde \rho_k = {\rm diag}(\rho_k)
\end{align}
and computing the mutual information
\begin{align}
    {\cal I}(\tilde A:\tilde B)
    = S(\tilde \rho_A) + S(\tilde \rho_B) - S(\tilde\rho_{AB})
\end{align}
This definition is physically appealing since it reduces to the
mutual information for a purely mixed state and is independent of basis due to the invariance of the trace under the unitary transformation of the density matrix.  Furthermore, ${\cal I}(\tilde A:\tilde B) \le {\cal I}(A:B)$, implying that any difference between the two is entirely
due to quantum nature of correlations. 
With this, we define $D(A:B)$ as the 
{\em degree of quantumness} between subsystems 
A and B,
\begin{align}
    D(A:B) = \cal{I}(A:B)-\cal{I}(\tilde A:\tilde B).
    \label{eq:disco}
\end{align}
In testing the approach, this generally
under-estimates the 
 quantum correlation and, hence, 
provides a {\em lower} bound on the quantum-ness. On the 
other hand, $D(A:B)$ 
is strictly bound from above
by ${\cal I}(A:B)$. 

In Fig. \ref{fig:4}, we present mutual information and the corresponding {\em degree of quantumness} across various parametric regimes. Initially, in the scenario where the two spins lack direct exchange coupling (Fig.\ref{fig:4}(b)), it is observed that quantum contributions predominantly account for nearly all the correlation between the two spins. This is unexpected, given that the interaction between the two spins is mediated solely through the shared environment. Evidently, in the context of uncorrelated environments, there is an absence of shared information. Conversely, in correlated environments, quantum correlations predominantly contribute to the mutual information, whereas in the anti-correlated scenario, classical contributions assume a dominant role.


\section{Discussion}
Synchronization has
been a rich and
diverse topic in both
quantum and classically
described systems for well over 350 years.
In most studies of this sort, the two oscillators are linked in some direct way via mechanical coupling as
in the case of a classical pendula or via exchange of quanta. 
Here, we show that
indirect coupling mediated by contact with a common stochastic environment
can also lead to quantum entanglement and synchronization even in the absence of
direct coupling. 
While coupling to a shared resource leads to 
correlation, that correlation may be quantum or
classical in origin.  We propose that
this can be quantified by subtracting the
maximal amount of classical information shared
between the two components
from their total mutual entropy, ${\cal I}(A:B)$,  and thus provides a lower bound for the
amount of quantum information shared between the two
components.  
Our analysis indicates that 
correlated noise leads to a high-degree
of ``quantum-ness'' in steady-state system, anti-correlated noise tends to result in more classical-like correlation.  Furthermore, 
when the direct coupling between the
spins is eliminated, nearly 100\% of
the correlation between the two qubits
can be attributed to quantum
contributions.


\begin{acknowledgments}
ERB acknowledges funding from the National Science Foundation (CHE-2404788) and the Robert A. Welch Foundation (E-1337). 
HL and CSA acknowledge funding from the Government of Canada (Canada Excellence Research Chair CERC-2022-00055) and the Courtois Institute, Facult\'e des arts et des sciences, Universit\'e de Montr\'eal (Chaire de Recherche de l'Institut Courtois).
The authors also acknowledge Prof. Phillip Shushkov (Indiana University) for discussions at a recent Telluride (TSRC) workshop that led to this investigation. 
\end{acknowledgments}

\section*{Data Availability Statement}
Data files and source codes used in this work are available on GitHub at \url{https://github.com/BittnerTheoryGroup/Quantum_Synchro-131}.

\section*{Author Contribution Statement}
ERB conceived of the idea. HL and ERB developed the correlated noise model. ERB and BT developed the theory and performed the computations.
ERB
supervised the findings of this work. 
All authors discussed the results and contributed to the final manuscript.

\section*{Competing Interests Statement}
The authors have no
competing interests in this work. 

\section*{Supplemental Information}
Movies showing phase locking behavior for model qubit dimers
are available on GitHub (\url{https://github.com/BittnerTheoryGroup/Quantum_Synchro-131}). 




\bibliography{References_local}

\newpage

\appendix
\section{Comments on Approximating the Discord.}
\begin{figure*}[b]
    \centering
    \subfigure[]{\includegraphics[width=0.3\linewidth]{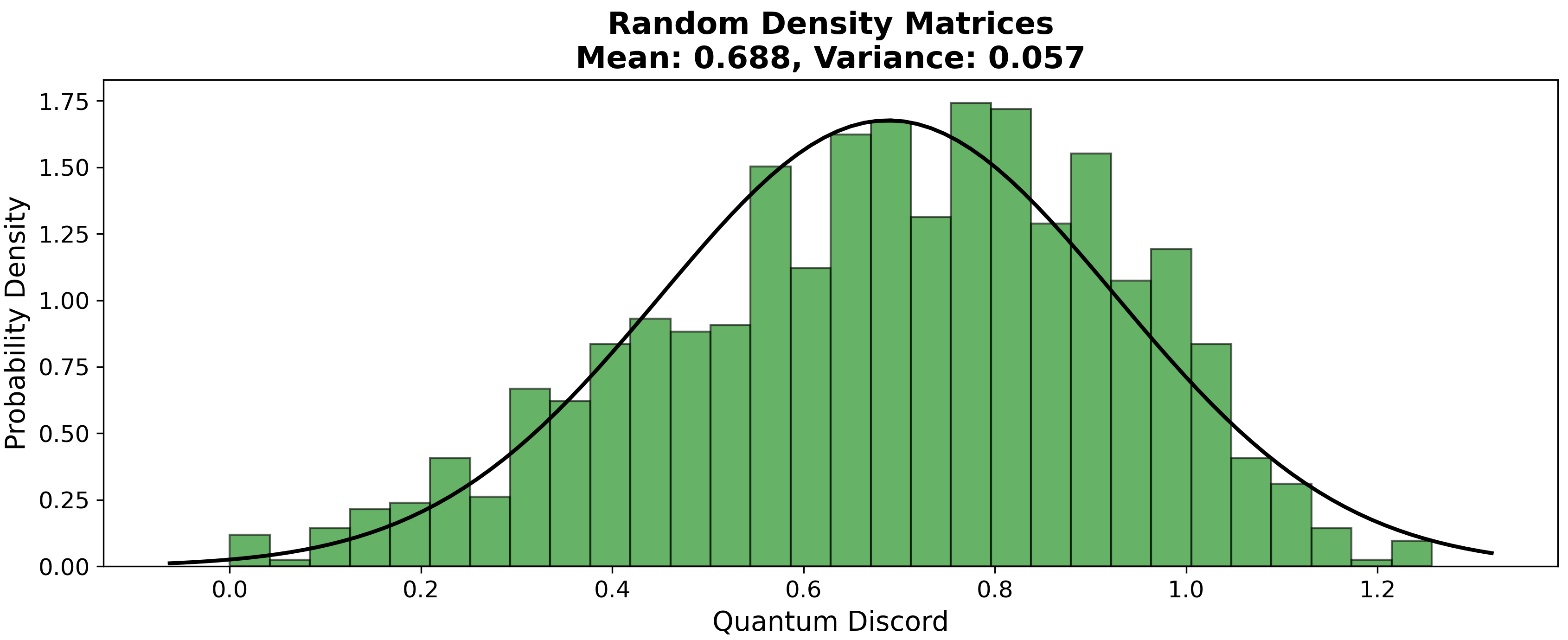}}
    \subfigure[]{\includegraphics[width=0.3\linewidth]{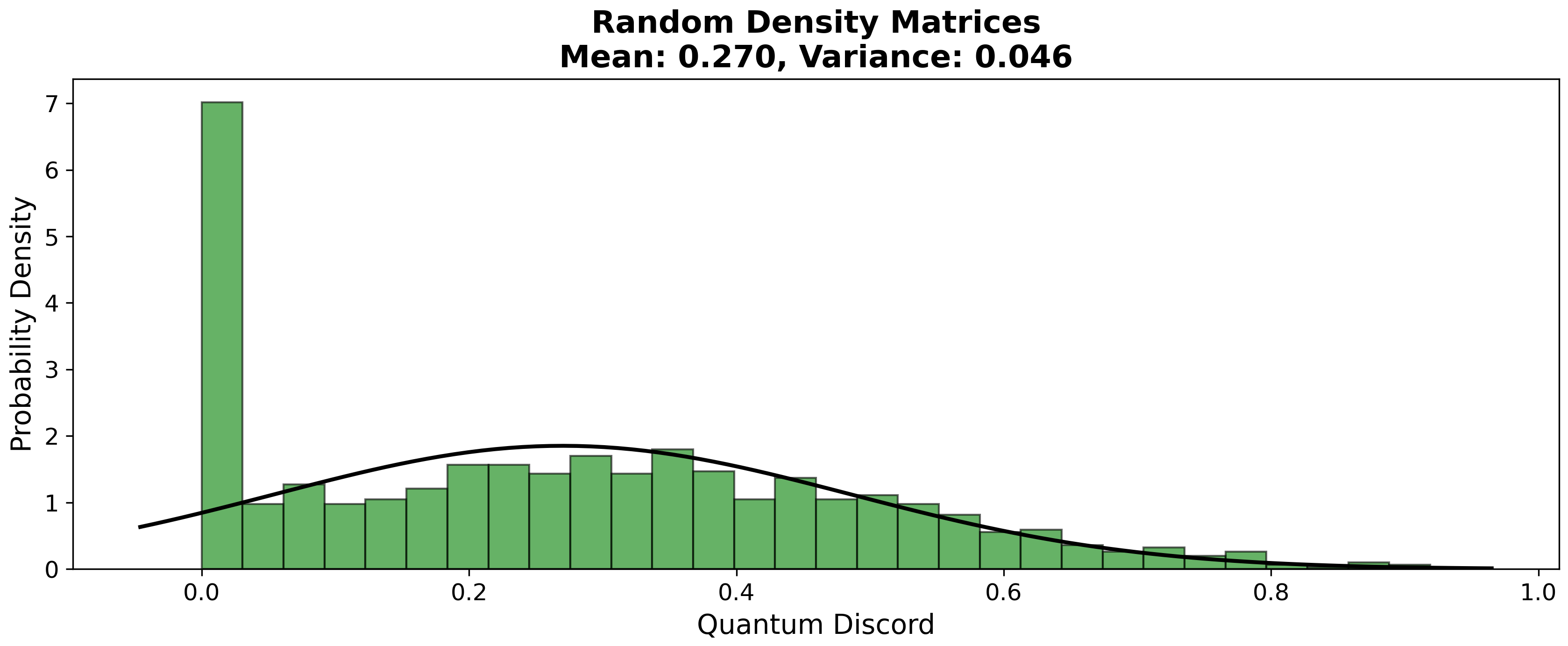}}
    \subfigure[]{\includegraphics[width=0.3\linewidth]{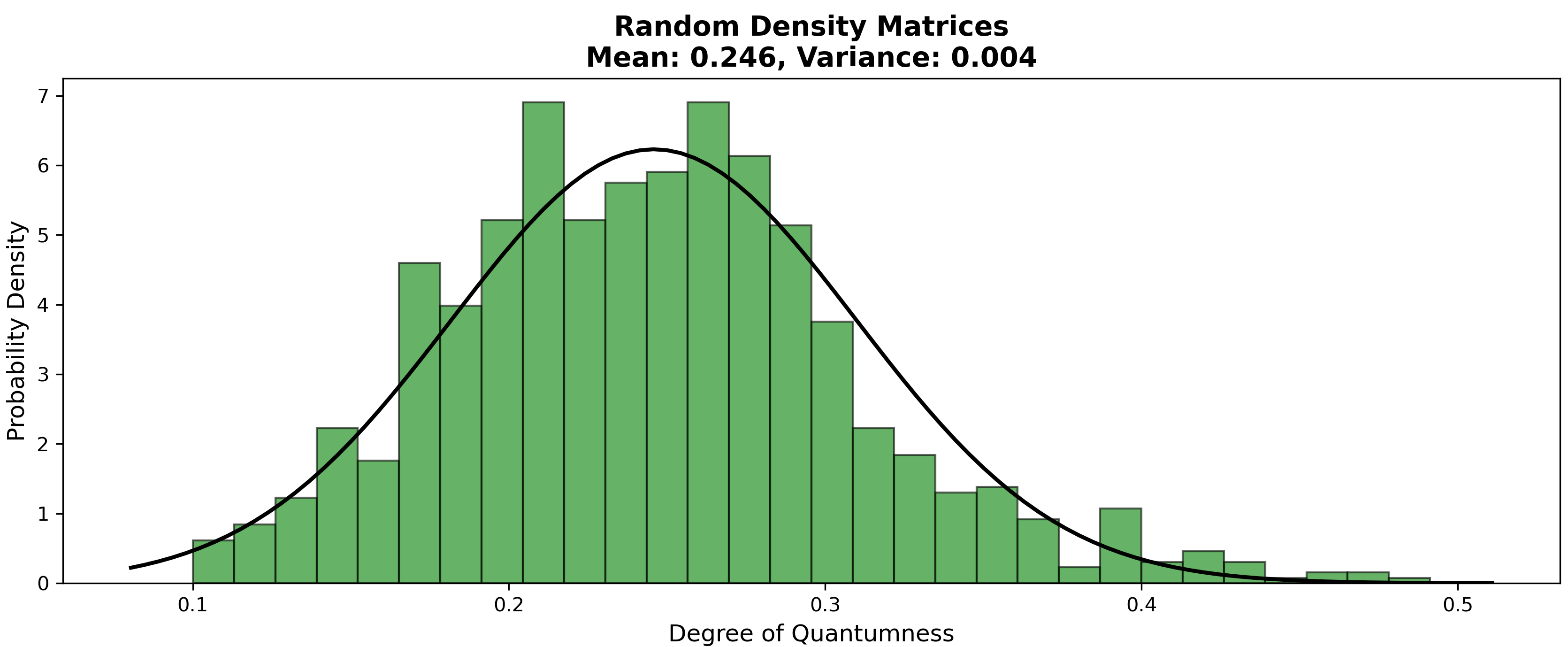}}
    \caption{The quantum discord for a thousand random rank-2 matrices (a) and the tight upper bound on the discord for a thousand random density matrices (including rank-2, 3 and 4)(b) for a two-qubit system can be exactly calculated using the method described by Galve et. al\cite{Galve_2011}. The {\em degree of quantumness} (c) calculated using the approximation used in Section \ref{syncdynamics} underestimates the amount of quantum discord.}
    \label{fig:5}
\end{figure*}
Despite having an exact expression to calculate the amount of true quantum correlations that exist between the subsystems, this calculation is non-trivial because it requires us to minimise over all possible sets of allowed measurements on one of the subsystems. This optimization is challenging because, given a general random unitary matrix, we can rotate this set of measurements to a completely new and equally valid set of measurements. 

For a system of qubits, several arguments can significantly reduce the cardinality of the set of allowed measurements and the complexity of the problem. The first result of Hamieh \textit{et al.} \cite{Hamieh2004} shows that the conditional entropy $S(\rho_A|k)$ is a concave function over the convex set of POVMs; hence, the set of allowed measurements or POVMs which correctly minimise the classical part of the correlations are extremal, which means that they cannot be constructed via a convex combination of other POVMs. Following this, D'Ariano \textit{et al.} \cite{D'Ariano_2005} showed that these extremal POVMs for qubits are rank-1 and the allowed set can only have two, three or four elements. Finally, Galve \textit{et al.} \cite{Galve_2011} showed that for a system of two qubits, the two-element set of POVMs corresponds to orthogonal measurements. This is based on the fact that a rank-1 POVM of two elements must satisfy positivity, $\Pi_{1,2}>0$, and normalization such that $\Pi_1 + \Pi_2= \mathbb{I}$. They invented a method to calculate the discord and provided numerical evidence that this calculation is exact (with an error around $10^{-4}$) for a system of two qubits described by a rank-2 density matrix. This method also provides a tight upper bound when the system is described by rank-3 and rank-4 density matrices. We show evidence for this in Fig.~\ref{fig:5} by calculating the discord for random density matrices describing a two-qubit system. 

For rank-2 density matrices in Fig.~\ref{fig:5}(a), these matrices 
have higher purity compared to general random matrices; hence, we see that the majority have discord between 0.75 and 1. Note that the quantum discord for any maximally entangled pure state is one and zero if the state is separable. For random density matrices (including rank-2, 3, and 4) in Fig.~\ref{fig:5}(b), when a system evolves under quantum dynamics and is also subject to some noisy environment, the output density matrix will most likely be mixed. Therefore, we see a tight upper bound as the quantum discord for maximally mixed density matrices is zero. It is important to note here that these methods were developed for systems in relatively small state spaces.  

The search for the most general method for identifying the optimal allowed measurement remains an active area of research. 
This motivated us to find an approximate and numerically efficient method to calculate the degree of quantumness that exists in the correlations of our final density matrix, described in Section \ref{syncdynamics}. We argue that by calculating the difference between the mutual information of our original density matrix $\rho_k$ and the mutual information of the modified density matrix $\tilde{\rho}_k = \mathrm{diag}(\rho_k)$, we can estimate the degree of quantum correlation in our quantum system. See Fig.~\ref{fig:5}(c). The intuition behind this approximation is that the diagonal elements of a density matrix contain all the classical information, i.e., the information that we get after local measurements. So when we ignore the off-diagonal elements, we consider only the classical part of the distribution and remove it from the total mutual information. This approximation provides an upper bound on the amount of classical correlations and a lower bound on the quantum correlations. One advantage of this approximation is that it can be extended to higher-spin systems and can be used as a quick estimation to extract the amount of quantum correlations.

\end{document}